# Novel Recursive Inclusion-Exclusion Technology Based on BAT and MPs for Heterogeneous-Arc Binary-State Network Reliability Problems

Wei-Chang Yeh
Integration and Collaboration Laboratory
Department of Industrial Engineering and Engineering Management
National Tsing Hua University
yeh@ieee.org
+886-986555381

*Abstract* — Current network applications, such as utility networks (gas, water, electricity, and 4G/5G), the Internet of Things (IoT), social networks, and supply chains, are all based on binary state networks. Reliability is one of the most commonly used tools for evaluating network performance, and the minimal path (MP) is a basic algorithm for calculating reliability. However, almost all existing algorithms assume that all undirected arcs are homogeneous; that is, the probability of an arc from nodes *a* to *b* is equal to that from nodes *b* to *a*. Therefore, based on MPs, the binary-addition-tree algorithm (BAT), and the inclusion-exclusion technique (IET), a novel recursive inclusion-exclusion technology algorithm known as recursive BAT-based IET (RIE) is proposed to solve the heterogeneous-arc binary-state network reliability problem to overcome the above obstacles in applications. The computational complexity of the proposed RIE is analyzed using an illustrative example. Finally, 11 benchmark problems are used to verify the performance of RIE.

## 1. INTRODUCTION

Various network models have been widely implemented in many different traditional and modern applications globally, such as power distribution networks [1, 2], water networks [3], gas networks [4], common networks [5], logistics networks [6], industry 4.0 networks [7], 5G/6G networks [8], the Internet of Things (IoT) [9], wireless sensor networks [10], and deep learning [11–15].

A binary-state network is a basic network, the components of which are binary; that is, either success or failure. Based on binary-state networks, multi-state networks extend the number of states from two to two more [16–18]. Among the multi-state networks, multi-commodity networks allow for more than one type of flow [19]. As an extension of multi-commodity networks, multi-distribution networks may have more than one distribution for each component [20]. Thus, the binary-state network is the origin of all network categories, and the advancement of algorithms in binary-state network problems is beneficial for all types of network problems [19–21].

Network reliability reflects the current status probability of a network. Hence, network reliability has been one of the most popular tools for indicating network performance for several decades. Traditional network reliability problems simply focus on calculating the network reliability without considering other constraints [22–29]. Owing to the rapid development of advanced technologies, real-life network reliability problems are increasingly focused on conditional network reliability to account for constraints such as *k*-out-of-*n* [30], budget constraints [31–32], limited transmission speed [33], limited memory capacity, and signal quality.

Conditional network reliability problems mainly determine the minimal paths (MPs) to satisfy predefined conditions in advance. An MP is a special path from nodes 1 to *n* such that the removal of any of its arcs results in no connection between nodes 1 and *n*. After determining the MPs, inclusion-exclusion technology (IET) is the most convenient technique for calculating the network reliability in terms of MPs among all algorithms, such as binary decision diagram, sum of disjoint product, and universal generating function. The two procedures described above for determining the MPs and calculating network reliability using MPs are, namely, NP-hard and #P-hard [48, 49].

A heterogeneous arc is a special arc with two directions, and each direction has its own state probability. Heterogeneous arcs appear in many applications; for example, the probability of a traffic jam occurring from location A to B is different from that from B to A. It is also common to replace a heterogeneous arc with arcs with two opposite directions. However, such a replacement results in the square of the runtime in calculating heterogeneous-arc networks.

IET is straightforward and plays a significant role in determining the union of different sets in many applications. IET adds and subtracts different intersections of sets, which are known as IET terms, to obtain the results. Hence, it is always necessary to obtain an efficient IET with fewer IET terms and fewer intersections in forming the IET terms, including the calculation of the heterogeneous-arc network reliability in terms of MPs directly, without the above replacement.

A new IET known as recursive BAT-based IET (RIE) is proposed to improve IET and solve heterogeneous-arc network reliability problems. Yeh [34] first proposed BAT to generate all possible binary vectors and applied it to binary-state network reliability problems to filter out all connected vectors for calculating network reliability without using MPs and minimal cuts (MCs).

To the best of our knowledge, no recursive IET for heterogeneous-arc network reliability problems or BAT for determining all MPs has been reported till date. Moreover, all known MP-based algorithms focus on homogeneous arcs, which have a limitation that both directions have the same reliability for each arc. Hence, the goal of this study is to develop RIE to integrate the BAT, recursive concept, and directed MPs to reduce the number of IET terms and interactions to obtain IET terms for solving the heterogeneous-arc network reliability efficiently.

The remainder of this paper is organized as follows: All acronyms, notations, nomenclature, and assumptions that are required in the proposed RIE are described in Section 2. A brief overview of MPs, IET, BAT, and heterogeneous arcs is provided in Section 3. The novelty of the proposed RIE is discussed in detail in Section 4. The pseudocode of the proposed RIE is explained and a demonstration is presented in Section 5. An experiment on the time complexity and a complete comparative experiment using 11 benchmark problems with QIE, which remains the best-known IET algorithm, are also presented in Section 5. Finally, Section 6 provides concluding remarks and discusses future work.

## 2. ACRONYMS, NOTATIONS, NOMENCLATURE, AND ASSUMPTIONS

All of the required acronyms, notations, nomenclature, and assumptions are provided in this section.

### 2.1 Acronyms

- MP: minimal path
- MC: minimal cut
- IET: inclusion–exclusion technology
- QIE: quick IET proposed in [35]
- BAT: binary-addition tree algorithm [34]
- RIE: Proposed recursive BAT-based IET

### 2.2 Notations

- $|\bullet|$: number of elements in set $\bullet$
- $\Pr(\bullet)$: success probability of $\bullet$
- $n$: number of nodes

$m$: number of arcs without considering arc directions

$p$: number of MPs

$V$: set of nodes $V = \{1, 2, \ldots, n\}$

$E$: set of arcs

$a_k$: $k^{\text{th}}$ arc in $E$

$e_{i,j}$: directed arc from nodes $i$ to $j$

$\varepsilon_{i,j}$: undirected arc between nodes $i$ and $j$

$m^*$: number of directed arcs without considering $e_{i,0}$ and $e_{n,i}$ for all nodes $i$

$\mathbf{D_b}$: The state distribution lists the functioning probability for each arc; e.g., $\mathbf{D_b} = \{\Pr(e_{1,2}) = \Pr(e_{1,3}) = \Pr(e_{2,3}) = \Pr(e_{2,4}) = \Pr(e_{3,4}) = 0.9, \Pr(e_{2,1}) = \Pr(e_{3,1}) = \Pr(e_{3,2}) = \Pr(e_{4,2}) = \Pr(e_{4,3}) = 0.8\}$ of the heterogenous-arc binary-state network in Figure 1.

$G(V, E)$: A graph with $V$, $E$, source node 1, and sink node $n$; for example, Figure 1 is a graph with $V = \{1, 2, 3, 4\}$, $E = \{e_{1,2}, e_{2,1}, e_{1,3}, e_{3,1}, e_{2,3}, e_{3,2}, e_{2,4}, e_{4,2}, e_{3,4}, e_{4,3}\}$, source node 1, and sink node 4.

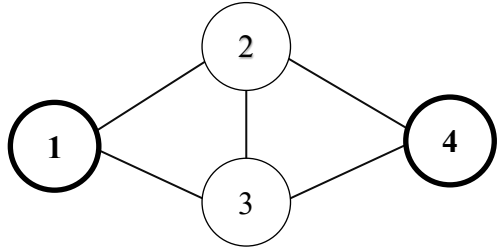

**Figure 1.** Example network.

$G(V, E, \mathbf{D_b})$: A binary-state network with $G(V, E)$ and $\mathbf{D_b}$; e.g., $G(V, E)$ is a binary-state network in Figure 1 after $\mathbf{D_b}$ is determined.

$R(G)$: reliability of network $G(V, E, \mathbf{D_b})$

$X$: binary vector

$X(a)$: coordinate related to arc $a$ in $X$ for $a \in E$

$Pr(X)$: 
$$Pr(X) = \sum_{X(a)=1} Pr(a)$$

$T_i$: The $i$th IET (intersection) term; e.g., $T_1 = \emptyset$, $T_2 = A$, $T_3 = B$, and $T_4 = A \cap B$ in $Pr(A \cap B) = Pr(A) + Pr(B) - Pr(A \cap B)$.

$S(T_i)$: The sign of IET term $Pr(T_i)$ such that $S(T_i) = 1$ and $-1$ if $T_i$ is the intersection of the odd and even numbers of elements, respectively; e.g., $S(T_1) = -1$, $S(T_2) = S(T_3) = 1$, and $S(T_4) = -1$ in $Pr(A \cap B) = Pr(A) + Pr(B) - Pr(A \cap B)$.

$N_\bullet$: number of IET terms obtained from algorithm $\bullet$

$T_\bullet$: seconds of runtimes obtained from algorithm $\bullet$

## 2.3 Nomenclature

Heterogeneous arc: An undirected arc can be separated into two opposite-direction arcs, and each directed arc has its own probability; for example, $Pr(e_{1,2}) = 0.9 \neq Pr(e_{2,1}) = 0.1$.

Reliability: The probability that nodes 1 and $n$ are connected by one path.

MP: A path set such that none of its proper subset arcs is an MP. For example, $\{e_{1,2}, e_{2,3}, e_{3,4}\}$ is an MP from nodes 1 and 4 in Figure 1.

IET term: Each independent intersection, including an empty set is known as an IET term; e.g., $\emptyset$, $Pr(A)$, $Pr(B)$, and $Pr(A \cap B)$ are four terms in $Pr(A \cap B) = Pr(A) + Pr(B) - Pr(A \cap B)$.

Augmented-state vectors: A new vector is obtained by extending the binary state to four states for each coordinate to reduce the number of arcs such that the states of $e_{i,j} = 0, 1, 2, 3$, for all $i < j$, are denoted as both $e_{i,j}$ and $e_{j,i}$ have failed; $e_{i,j}$ is working and $e_{j,i}$ has failed; $e_{i,j}$ has failed and $e_{j,i}$ is working; and both $e_{i,j}$ and $e_{j,I}$ are working, respectively. Note that $e_{i,j}$ is not considered in the augmented-state vectors for all $i > j$.

**2.4 Assumptions**

1. Each node is perfectly reliable and connected.

2. Each arc state is either functioning or has failed.

3. No parallel arcs or loops exist.

4. Each arc state probability is statistically independent with a predetermined distribution.

**3. OVERVIEW OF MP, IET, BAT, AND HETEROGENEOUS ARCS**

Prior to presenting the proposed RIE for calculating the heterogeneous-arc binary-state network reliability problem, the fundamental concepts of the proposed RIE, including MPs, IET, BAT, and heterogeneous arcs, are briefly introduced.

**3.1 MP and MP-Based Algorithms**

The most popular network-based tools for binary-state networks relate to either MCs or MPs. An MP is a special path that is an arc subset for connecting a sequence of nodes between the source and sink nodes. An MP is not an MP if any arc is removed from it, and this special characteristic is important in verifying whether a path is an MP.

The majority MP generation methods are based on heuristic algorithms [36], universal generating function methodologies [21, 25, 37], and implicit enumeration algorithms [38].

MPs play an important role in MP-based algorithms to calculate the exact reliability of general binary-state networks. However, each MP-based algorithm is an indirect algorithm and other methods, such as IET or sum-of-disjoint products [39–41], are required to calculate the final reliability in terms of determining the MPs. Unfortunately, the above two steps, namely, searching for all MPs/MCs and applying these methods in terms of MPs/MCs for calculating the reliability, are both NP-hard problems [48, 49].

## 3.2 IET

Despite its limited network reliability, IET plays an important role in counting the union of possible non-disjoint sets in terms of the intersections of all subsets for numerous applications and research areas. Hence, owing to its simplicity, many variants of IET have been proposed, and many improvements have been implemented to enhance its efficiency.

Let $P = \{P_1, P_2, \ldots, P_p\}$ represent a collection of finite sets, that is, the MPs in this study; and let $I_k$ represent all possible intersections of $k$ sets that are selected without replacement from $P$. Each intersection in $I_k$ is known as an IET term [35, 39, 40, 41]. Then,

$$\Pr\left(\bigcup_{i=1}^{p} P_i\right) = \sum_{k=1}^{p} (-1)^{k+1} \Pr(\bigcap_{i \in I_k} P_i). \tag{1}$$

Based on the aforementioned fundamental concept of set theory, IET is the most straightforward and convenient technique that has been discussed at length in the literature for calculating the exact reliability and reliability bounds of all network types.

## 3.3 BAT

BAT was first developed by Yeh [34] to generate all the required binary vectors for various problems with binary decision variables [34, 44–47]. In calculating the binary-state reliability, it outperforms most traditional algorithms such as binary decision diagram, depth first search, breadth first search, and universal generating function [34, 44].

In BAT, only one binary vector $X$ exists, which is initialized to be vector zero, and it is updated repeatedly to generate all possible vectors or solutions for the problems based on the following simple rule [34]:

If $a_i = 0$ and $a_j = 1$, let $a_i = 1$ and $a_j = 0$, where for $j = 1, 2, …, (i–1) < (m–1)$. Otherwise, halt, and all vectors are identified. For example, the vectors (0, 1, 0) and (1, 0, 0) are updated to (1, 1, 0) and (0, 1, 0), respectively. Furthermore, the complete process halts after the vector with no zero-value coordinates is identified, which is (1, 1, 1).

The details of the BAT pseudo-code for carrying out the above simple rule are as follows [34, 51]:

**Algorithm for finding All *m*-tuple Binary Vectors**

**Input:**  *m*.

**Output:**  All *m*-tuple binary vectors.

**STEP B0.** Let $X = \mathbf{0}$ and $i = 1$.

**STEP B1.** If $X(a_i) = 0$, let $X(a_i) = 1$, $i = 1$, $X$ be updated, and return STEP B1.

**STEP B2.** If $i = m$, halt.

**STEP B3.** Let $X(a_i) = 0$, $i = i + 1$, and return STEP B1.

Only four statements appear in the BAT pseudo-code; only one vector $X$ is updated repeatedly, and it is only verified whether the current coordinate is zero. Hence, BAT is simple to code, easy to understand, memory-friendly, and efficient in operation.

The updated $X$ in the above pseudo-code is used once without the need to store it; this implies BAT is you-only-look-once (YOLO). Hence, BAT can solve various problems if there is no time limit, without running out of memory [34]. Note that the major problem for these recursive methods, such as recursive SDP, is that they always lack memory if the problem size is medium [34, 35].

Moreover, in STEP B1, after obtaining a new $X$, it can be verified whether $X$ is feasible or optimum, depending on the practical problem. For example, $X$ is feasible when it is connected at a binary network. Therefore, BAT is very convenient and has been applied to many research areas, such as resilience assessment [45], the spread of wildfires [46], the propagation of computer viruses [47], and the reliability of various networks [34, 44].

**3.4 Heterogeneous Arcs**

A heterogeneous arc has two distinctive success probabilities for the same arc in two different directions. In contrast, if the success probabilities of the two opposite-direction arcs are equal, the arc is homogeneous. For example, the undirected arc $\varepsilon_{1,2}$ is a heterogeneous arc or a homogeneous arc if $\Pr(e_{1,2}) \neq \Pr(e_{2,1})$ and $\Pr(e_{1,2}) = \Pr(e_{2,1})$, respectively.

In homogeneous-arc binary-state networks, the directions of each arc can be ignored when calculating the reliability. However, in general, each undirected heterogeneous arc is replaced with two directed arcs in binary-state network reliability problems [34], so that no heterogeneous arcs are adjacent to nodes 1 or $n$. For example, the graph after replacing undirected arcs with directed arcs in Figure 2 is depicted in Figure 2.

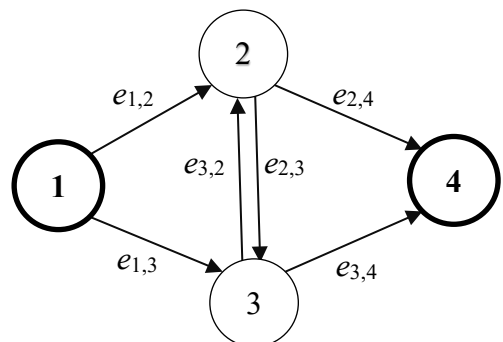

**Figure 2.** Graph after replacing undirected arcs with directed arcs in Figure 1.

According to Figure 2, the time complexity of these algorithms for heterogeneous-arc network reliability problems is the square of that of homogeneous-arc network reliability problems. Thus, it is necessary to derive a simple algorithm without the need to transfer each undirected heterogeneous arc to two directed arcs to reduce the computational burden of the network reliability problem.

## 4. MAJOR NOVELTIES OF PROPOSED ALGORITHM

Several major components of the proposed RIE are included to calculate the heterogeneous-arc binary-state network reliability: 1) transferring the undirected MPs to directed MPs without replacing each undirected arc with two directed arcs, 2) implementing the BAT-based IET in terms of directed MPs using the proposed augmented -state vectors to calculate the network reliability, and 3) removing duplicated complete terms in the RIE.

These three aspects are discussed in detail in the following subsections.

### 4.1 Directed MPs

Most MP-based algorithms can only identify all undirected MPs for networks in which all arcs are homogenous. Only a few MP algorithms can include only directed MPs after transferring each undirected arc to two directed arcs in binary-state networks with only heterogeneous arcs [34].

The first step of the proposed algorithm is to identify all MPs. However, all MPs are initially undirected, as in most MP-based algorithms. To reduce the runtime, a simple algorithm is proposed to determine the directed MPs in terms of undirected MPs without the need to transfer each undirected arc into two directed arcs.

The pseudo-code for the proposed algorithm is presented as follows:

**Algorithm for determining directed MPs from undirected MPs**

**Input:**     A graph $G(V, E)$ with source node 1, sink node $n$, and all undirected MPs $Q_1, Q_2, \ldots, Q_p$ is the input.

**Output:**    All directed MPs $P_1, P_2, \ldots, P_p$.

**STEP M0.** Let $i = j = l = 1$ and $P_i = \varnothing$ for all $i = 1, 2, \ldots, p$.

**STEP M1.** Let $a$ be the $l$th undirected arc from the endpoint node $j$ to its another endpoint node $k$ in $Q_i$.

**STEP M2.** Let $P_i = P_i \cup \{e_{j,k}\}$.

**STEP M3.** If $l < |Q_i|$, let $l = l + 1$ and go to STEP M1.

**STEP M4.** If $i < p$, let $i = i + 1$, $j = l = 1$, and go to STEP M1. Otherwise, halt.

The proposed algorithm is based on the characteristics of the MP, where no arc is redundant in any MP. Hence, the direction of the directed MP is always from node 1 to $n$, and STEP M1 is based on these important characteristics.

As at most $n$ arcs exist in each MP, the pseudo-code above requires $O(|V|)$ to determine the direction of each arc in any MP in the worst case. Hence, the time complexity to change all undirected MPs into all directed MPs is only $O(np\pi)$, where $O(\pi)$ is the time complexity for finding all undirected MPs.

The number of arcs is doubled after changing each undirected arc into two directed arcs. Hence, the pseudo-code is more efficient and simple compared to algorithms for determining directed arcs with a time complexity of $O(np^2\pi)$ by changing each undirected arc into two directed arcs.

For example, in Figure 1, $Q_4 = \{a_2, a_3, a_4\}$ is an undirected MP. The complete process after implementing the pseudo-code above for transferring $Q_4$ to directed MP $P_4$ is presented in Table 1.

**Table 1.** Directed MP of $Q_4 = \{a_2, a_3, a_4\}$ in Figure 1.

| $i$ | $j$ | $k$ | $a$ | $P_4$ |
|---|---|---|---|---|
| 1 | 1 | 3 | $a_2$ | $\{e_{1,3}\}$ |
| 2 | 3 | 2 | $a_3$ | $\{e_{1,3}, e_{3,2}\}$ |
| 3 | 2 | 4 | $a_4$ | $\{e_{1,3}, e_{3,2}, e_{2,4}\}$ |

**4.2 BAT-Based IET**

To obtain each IET term, which is an intersection of some MPs, easily, systematically, and directly, the proposed BAT-based IET obtains all IET terms based on the vector that is obtained from the BAT.

The major difference between the proposed BAT-based IET and traditional BAT is that the $i$th coordinate in each vector in the BAT-based IET indicates whether the $i$th directed MP is used in the IET terms. However, the $i$th coordinate in each vector is the state of the arc in traditional BAT.

For example, four MPs are shown in Figure 1. The binary vector of the BAT that is obtained from the proposed BAT-based IET is a 4-tuple, as indicated in the second column of Table 2. Each 4-tuple binary vector denotes an IET term, which is the intersection of all MPs in $\{P_i \mid X(i) = 1 \text{ for all } i\}$; for example, $X_7 = (0, 1, 1, 0)$ represents the IET term of $P_2 \cap P_3$, as indicated in the third column of Table 2, and the elements in $P_2 \cap P_3$ are displayed in the fifth column of Table 2.

**Table 2.** The BAT and the IET terms

| $i$ | $X_i$ | $T$ | Elements in $T$ | $S(T)$ | $\Pr(T)$ | $R$ |
|---|---|---|---|---|---|---|
| 1 | (0, 0, 0, 0) | | | | | |
| 2 | (1, 0, 0, 0) | $P_1$ | $e_{1,2}, e_{2,4}$ | + | 0.81 | 0.81 |
| 3 | (0, 1, 0, 0) | $P_2$ | $e_{1,2}, e_{2,3}, e_{3,4}$ | + | 0.729 | 1.539 |
| 4 | (1, 1, 0, 0) | $P_1 \cap P_2$ | $e_{1,2}, e_{2,4}, e_{2,3}, e_{3,4}$ | – | 0.6561 | 0.8829 |
| 5 | (0, 0, 1, 0) | $P_3$ | $e_{1,3}, e_{3,4}$ | + | 0.81 | 1.6929 |
| 6 | (1, 0, 1, 0) | $P_1 \cap P_3$ | $e_{1,2}, e_{2,4}, e_{1,3}, e_{3,4}$ | – | 0.6561 | 1.0368 |
| 7 | (0, 1, 1, 0) | $P_2 \cap P_3$ | $e_{1,2}, e_{2,3}, e_{3,4}, e_{1,3}$ | – | 0.6561 | 0.3807 |
| 8 | (1, 1, 1, 0) | $P_1 \cap P_2 \cap P_3$ | $e_{1,2}, e_{2,4}, e_{2,3}, e_{3,4}, e_{1,3}$ | + | 0.59049 | 0.97119 |
| 9 | (0, 0, 0, 1) | $P_4$ | $e_{1,3}, e_{3,2}, e_{2,4}$ | + | 0.648 | 1.61919 |
| 10 | (1, 0, 0, 1) | $P_1 \cap P_4$ | $e_{1,2}, e_{2,4}, e_{1,3}, e_{3,2}$ | – | 0.5832 | 1.03599 |
| 11 | (0, 1, 0, 1) | $P_2 \cap P_4$ | $e_{1,2}, e_{2,3}, e_{3,4}, e_{1,3}, e_{3,2}, e_{2,4}$ | – | 0.472392 | |
| 12 | (1, 1, 0, 1) | $P_1 \cap P_2 \cap P_4$ | $e_{1,2}, e_{2,4}, e_{2,3}, e_{3,4}, e_{1,3}, e_{3,2}$ | + | | |
| 13 | (0, 0, 1, 1) | $P_3 \cap P_4$ | $e_{1,3}, e_{3,4}, e_{3,2}, e_{2,4}$ | – | 0.5832 | 0.45279 |
| 14 | (1, 0, 1, 1) | $P_1 \cap P_3 \cap P_4$ | $e_{1,2}, e_{2,4}, e_{1,3}, e_{3,4}, e_{3,2}$ | + | 0.52488 | 0.97767 |
| 15 | (0, 1, 1, 1) | $P_2 \cap P_3 \cap P_4$ | $e_{1,2}, e_{2,3}, e_{3,4}, e_{1,3}, e_{3,2}, e_{2,4}$ | + | | |
| 16 | (1, 1, 1, 1) | $P_1 \cap P_2 \cap P_3 \cap P_4$ | $e_{1,2}, e_{2,4}, e_{2,3}, e_{3,4}, e_{1,3}, e_{3,2}$ | – | | |

In Eq. (1), the sign of each term is based on the number of MPs that are used in the intersection, such that it is positive if the number of MPs is an odd number; otherwise, it is negative, as indicated in the fourth column of Table 2. The probability of each available IET term is presented in the sixth column of Table 2. The details of the final two columns are presented in Section 4.3.

The maximum number of arcs in each MP is $n$ at most. Similarly, by replacing each IET term with two IET terms based on the BAT, the number of intersections in STEP L1 and the probability calculation of MPs in STEP L2 can be reduced from $O(2mp2^p)$ to $O(np2^p)$ and from $O(2m2^p)$ to $O(n2^p)$, respectively.

**4.3 Recursive Concept**

A reduction in the number of intersections can also decrease the time complexity of the IET [35]. A new concept known as the recursive concept, which replaces the current IET term with two IET terms based on BAT, is proposed to accomplish the above goal.

These MPs intersect to determine the term $T_i$ from $X_i$ in BAT, as noted in Section 4.2. From the BAT-IET, we obtain $X_j = X_k + X_i$, i.e., $T_j = T_k \cap T_i$, where $X_j(a_i) = 1$ and $X_j(a_i) = 0$ for all $l = (i+1), (i+2), \ldots, p$. For example, $X_8 = (1, 1, 1, 0) = (1, 1, 0, 0) + (0, 0, 1, 0) = X_4 + X_3$ and $T_8 = T_4 \cap T_3 = \{e_{1,2}, e_{2,4}, e_{2,3}, e_{3,4}, e_{1,3}\}$ according to Table 2. Hence, based on the above simple concept, the pseudo-code of the proposed recursive concept is presented as follows:

**Algorithm for proposed recursive BAT-IE**

**Input:** All directed MPs and terms $T_k$ for $k = 1, 2, \ldots, (i-1)$ are treated as input.

**Output:** $R(G)$.

**STEP R0.** Let $i = 1, j = 3, T_1 = \emptyset, T_2 = P_1, S(T_1) = -1, S(T_2) = 1, \Pr(T_1) = 1, \Pr(T_2) = \prod_{T_2(a)=1} \Pr(a), p^* = 2,$ and $R = 0$.

**STEP R1.** Let $k = 1$.

**STEP R2.** Let $T_j = T_k \cap T_i$ and $\Pr(T_j) = \Pr(T_k) \times \prod_{T_k(a)=0 \text{ and } P_i(a)=1} \Pr(a)$.

**STEP R3.** If $S(T_k) = -1$, let $S(T_j) = 1$, $R = R + \Pr(T_j)$, and go to STEP R5.

**STEP R4.** Let $S(T_j) = -1$, $R = R - \Pr(T_j)$.

**STEP R5.** If $k < p^*$, let $k = k + 1$, $j = j + 1$, and go to STEP R2.

**STEP R6.** If $i < p$, let $p^* = j$, $i = i + 1$, and go to STEP R1. Otherwise, $R = R(G)$ and halt.

From the above, we obtain the results shown in Table 3, which show all the results in Table 2 in a recursive manner. Following the implementation of the recursive concept, the number of intersections in each IET term is reduced from $(p-1)$ to 1. For example, in Table 3, $T_8 = T_4 \cap T_3 = \{e_{1,2}, e_{2,4}, e_{2,3}, e_{3,4}\} \cap \{e_{1,2}, e_{2,4}, e_{2,3}, e_{3,4}, e_{1,3}\}$ takes one intersection only. Moreover, the number of probability calculations is reduced; for example, $\Pr(T_8) = \Pr(T_4) \times \sum_{\varepsilon_k \in (T_1 - T_6)} \Pr(\varepsilon_k) = 0.6561 \times \Pr(\{e_{1,3}\}) = 0.59049$.

Therefore, the time complexity for the recursive BAT-IET is $O(n2^p)$, which is $(1/p)$ of that for the BAT-IET without using the recursive concept.

**Table 3.** The BAT and the IET terms

| $i$ | $X_i$ $T_i$ | $X_i$ $T_i$ | $X_i$ $T_i$ | $X_i$ $T_i$ |
|---|---|---|---|---|
| 1 | (0) ∅ | (0) ∅ | (0) ∅ | (0) ∅ |
| 2 | (1) $\{e_{1,2},e_{2,4}\}$ | (1) $\{e_{1,2},e_{2,4}\}$ | (1) $\{e_{1,2},e_{2,4}\}$ | (1) $\{e_{1,2},e_{2,4}\}$ |
| 3 | | (0,**1**) $\{e_{1,2},e_{2,3},e_{3,4}\}$ | (0,1) $\{e_{1,2},e_{2,3},e_{3,4}\}$ | (0,1) $\{e_{1,2},e_{2,3},e_{3,4}\}$ |
| 4 | | (1,**1**) $\{e_{1,2},e_{2,4},e_{2,3},e_{3,4}\}$ | (1,1) $\{e_{1,2},e_{2,4},e_{2,3},e_{3,4}\}$ | (1,1) $\{e_{1,2},e_{2,4},e_{2,3},e_{3,4}\}$ |
| 5 | | | (0,0,**1**) $\{e_{1,3},e_{3,4}\}$ | (0,0,1) $\{e_{1,3},e_{3,4}\}$ |
| 6 | | | (1,0,**1**) $\{e_{1,2},e_{2,4},e_{1,3},e_{3,4}\}$ | (1,0,1) $\{e_{1,2},e_{2,4},e_{1,3},e_{3,4}\}$ |
| 7 | | | (0,1,**1**) $\{e_{1,2},e_{2,3},e_{3,4},e_{1,3}\}$ | (0,1,1) $\{e_{1,2},e_{2,3},e_{3,4},e_{1,3}\}$ |
| 8 | | | (1,1,**1**) $\{e_{1,2},e_{2,4},e_{2,3},e_{3,4},e_{1,3}\}$ | (1,1,1) $\{e_{1,2},e_{2,4},e_{2,3},e_{3,4},e_{1,3}\}$ |
| 9 | | | | (0,0,0,**1**) $\{e_{1,3},e_{3,2},e_{2,4}\}$ |
| 10 | | | | (1,0,0,**1**) $\{e_{1,2},e_{2,4},e_{1,3},e_{3,2}\}$ |
| 11 | | | | (0,1,0,**1**) $\{e_{1,2},e_{2,3},e_{3,4},e_{1,3},e_{3,2},e_{2,4}\}$ |
| 12 | | | | (1,1,0,**1**) $\{e_{1,2},e_{2,4},e_{2,3},e_{3,4},e_{1,3},e_{3,2}\}$ |
| 13 | | | | (0,0,1,**1**) $\{e_{1,3},e_{3,4},e_{3,2},e_{2,4}\}$ |
| 14 | | | | (1,0,1,**1**) $\{e_{1,2},e_{2,4},e_{1,3},e_{3,4},e_{3,2}\}$ |
| 15 | | | | (0,1,1,**1**) $\{e_{1,2},e_{2,3},e_{3,4},e_{1,3},e_{3,2},e_{2,4}\}$ |

### 4.4 Augmented-State Vectors

The time complexity that is required to obtain the intersection of each IET term is $O(2mp)$ if each undirected arc is replaced with two directed arcs [34, 35]. Note that these directed arcs to node 1 and from node $n$ can all be deleted because it is impossible to use these arcs; for example, $e_{2,1}$, $e_{3,1}$, $e_{4,2}$, and $e_{4,3}$ in Figure 1 can be removed [34, 35].

In Figure 1, let $V(U) = (v_{1,2}, v_{1,3}, v_{2,3}, v_{2,4}, v_{3,4}, v_{3,2})$ be the vectors corresponding to set $U$ such that $V(U(a_i)) = v_i = 0$ or 1 if $a_i \in U$ or $a_i \notin U$, respectively; for example, $V(T_1) = (1, 0, 0, 1, 0, 0)$, $V(P_4) = (0, 1, 0, 1, 0, 1)$, and $V(T_{10}) = (1, 1, 0, 1, 0, 1)$. Hence, six comparisons are required to obtain $V(T_{10}) = V(T_1 \cap P_4) = (1, 0, 0, 1, 0, 0)$ because six directed arcs exist.

To reduce the number of comparisons between two vectors from $m^*$ to $m$, let $V^*(U(a_{i,j})) = v_{i,j} = 0, 1, 2,$ or 3 if $(a_{i,j} \notin U)$, $(a_{i,j} \in U)$ and $(a_{j,i} \notin U)$, $(a_{i,j} \notin U)$ and $(a_{j,i} \in U)$, or $(a_{j,i} \in U)$ and $(a_{i,j} \in U)$, respectively. Correspondingly, $\Pr(V^*(U(a_{i,j})) = 1$, $\Pr(a_{i,j})$, $\Pr(a_{j,i})$, and $\Pr(a_{i,j}) \times \Pr(a_{j,i})$ if $(a_{i,j} \notin U)$, $(a_{i,j} \in U)$ and $(a_{j,i} \notin U)$, $(a_{i,j} \notin U)$ and $(a_{j,i} \in U)$, or $(a_{j,i} \in U)$ and $(a_{i,j} \in U)$, respectively.

If $T_j = T_k \cap P_i$, we obtain $V^*(T_j) = V^*(T_k \cap P_i)$. The values of $V^*(T_j(a)) = V^*(T_k \cap P_i)(a)$ and $\Pr(V^*(T_k \cap P_i))$ can be calculated using the following two equations:

$$V^*(T_j(a)) = \begin{cases} V^*(T_k(a)) & \text{if } V^*(T_k(a)) = V^*(P_i(a)) \\ V^*(T_k(a)) + V^*(P_i(a)) & \text{otherwise} \end{cases}, \qquad (2)$$

$$\Pr(V^*(T_j)) = \Pr(V^*(T_k)) \times \prod V^*(P_i(a)) \quad \text{for all } a \text{ with } V^*(T_j(a)) \neq V^*(P_i(a)). \quad (3)$$

The pseudo-code that is used to calculate Eqs. (2) and (3) is as follows:

**Algorithm for calculating** $V^*(T_k \cap P_i)$ **and** $\Pr(V^*(T_k \cap P_i))$

**Input:** $T_k$ and $P_i$.

**Output:** $V^*(T_k \cap P_i)$ and $\Pr(V^*(T_k \cap P_i))$.

**STEP N0.** Let $l = R = 1$ and $V^*(T_j(a)) = V^*(T_k(a))$ for all $a$.

**STEP N1.** If $V^*(T_k(a_l)) = 0, 1, 2,$ and $3$, go to STEPs N2, N3, N4, and N5, respectively.

**STEP N2.** Let $V^*(T_j(a_l)) = V^*(P_i(a_l))$, $R = R \times \Pr(P_i(a_l))$, and go to STEP N5.

**STEP N3.** Let $V^*(T_j(a_l)) = 3$ and $R = R \times \Pr(P_k(a_l))$ if $V^*(P_k(a_l)) = 2$. Go to STEP N5.

**STEP N4.** Let $V^*(T_j(a_l)) = 3$ and $R = R \times \Pr(P_k(a_l))$ if $V^*(P_k(a_l)) = 1$. Go to STEP N5.

**STEP N5.** If $l < m^*$, let $l = l + 1$ and go to STEP N1. Otherwise, halt and let $\Pr(T_k \cap P_i) = R$.

For the example in the above, let $V^*(U) = (v_{1,2}, v_{1,3}, v_{2,3}, v_{2,4}, v_{3,4})$. Because $T_{10} = T_1 \cap P_4$, $V^*(T_1) = V^*(P_1) = (1, 0, 0, 1, 0)$, and $V^*(P_4) = (0, 1, 2, 1, 0)$, we have $V(T_{10}) = V^*(P_1) \oplus V^*(P_4) = (1+0, 0+1, 0+2, 1, 0) = (1, 1, 2, 1, 0)$ and $\Pr(V(T_{10})) = \Pr(1, 1, 2, 1, 0) = \Pr(V^*(T_1)) \times [\Pr(v_{1,3}=1) \times \Pr(v_{2,3}=2)] = 0.81 \times 0.9 \times 0.8 = 0.5832$. Note that the number of comparisons and multiplications are reduced from 6 to 5 after using the proposed concept of the new art states. Note that the number of comparisons and multiplications is reduced from six to five when the proposed concept of the new arc states is used.

Hence, a higher number of coordinates results in more time being saved when this concept is used. Table 4 lists all values of $V^* = (v_{1,2}, v_{1,3}, v_{2,3}, v_{2,4}, v_{3,4})$ for all IET terms when using the concept of the new arc states in Figure 1.

**Table 4.** The values of $V^* = (v_{1,2}, v_{1,3}, v_{2,3}, v_{2,4}, v_{3,4})$.

| $i$ | $T_i$ | $T_i$ | $V^*(T_i)$ |
|---|---|---|---|
| 1 | ∅ | ∅ | (0, 0, 0, 0, 0) |
| 2 | $P_1$ | $\{e_{1,2}, e_{2,4}\}$ | (1, 0, 0, 1, 0) |
| 3 | $P_2$ | $\{e_{1,2}, e_{2,3}, e_{3,4}\}$ | (1, 0, 1, 0, 1) |
| 4 | $P_1 \cap P_2$ | $\{e_{1,2}, e_{2,4}, e_{2,3}, e_{3,4}\}$ | (1, 0, 1, 1, 1) |
| 5 | $P_3$ | $\{e_{1,3}, e_{3,4}\}$ | (0, 1, 0, 0, 1) |
| 6 | $P_1 \cap P_3$ | $\{e_{1,2}, e_{2,4}, e_{1,3}, e_{3,4}\}$ | (1, 1, 0, 1, 1) |
| 7 | $P_2 \cap P_3$ | $\{e_{1,2}, e_{2,3}, e_{3,4}, e_{1,3}\}$ | (1, 1, 1, 0, 1) |
| 8 | $P_1 \cap P_2 \cap P_3$ | $\{e_{1,2}, e_{2,4}, e_{2,3}, e_{3,4}, e_{1,3}\}$ | (1, 1, 1, 1, 1) |
| 9 | $P_4$ | $\{e_{1,3}, e_{3,2}, e_{2,4}\}$ | (0, 1, 2, 1, 0) |
| 10 | $P_1 \cap P_4$ | $\{e_{1,2}, e_{2,4}, e_{1,3}, e_{3,2}\}$ | (1, 1, 2, 1, 0) |
| 11 | $P_2 \cap P_4$ | $\{e_{1,2}, e_{2,3}, e_{3,4}, e_{1,3}, e_{3,2}, e_{2,4}\}$ | (1, 1, 3, 1, 1) |
| 12 | $P_1 \cap P_2 \cap P_4$ | $\{e_{1,2}, e_{2,4}, e_{2,3}, e_{3,4}, e_{1,3}, e_{3,2}\}$ | (1, 1, 3, 1, 1) |
| 13 | $P_3 \cap P_4$ | $\{e_{1,3}, e_{3,4}, e_{3,2}, e_{2,4}\}$ | (0, 1, 2, 1, 1) |
| 14 | $P_1 \cap P_3 \cap P_4$ | $\{e_{1,2}, e_{2,4}, e_{1,3}, e_{3,4}, e_{3,2}\}$ | (1, 1, 2, 1, 1) |
| 15 | $P_2 \cap P_3 \cap P_4$ | $\{e_{1,2}, e_{2,3}, e_{3,4}, e_{1,3}, e_{3,2}, e_{2,4}\}$ | (1, 1, 3, 1, 1) |
| 16 | $P_1 \cap P_2 \cap P_3 \cap P_4$ | $\{e_{1,2}, e_{2,4}, e_{2,3}, e_{3,4}, e_{1,3}, e_{3,2}\}$ | (1, 1, 3, 1, 1) |

**4.5 Complete Terms**

The proposed recursive concept can only reduce the number of intersections and probability calculations. If the number of terms can be reduced, the efficiency of the IET can be directly increased.

The term with all arcs is known as the complete term; for example, $T_{11} = T_3 \cap T_4$ in Table 4. More than one complete term may exist; for example, $T_{11} = T_{12} = T_{15} = T_{16}$ in Table 4. A simple method for reducing the number of terms is the detection and removal of duplicated complete terms.

Let $X_i$ be the vector corresponding to the complete term $T_i$, and $0(X_i) = \{ k \mid X_i(k) = 0 \text{ for all } k \}$. Any IET term $T_h$ is also a complete term if $X_h(k) = X_i(k)$ for all $k \notin I_i$. The number of such vectors is $2^{|0(X_i)|}$, which is an even number if $0(X_i) \neq \emptyset$ and 1 if $0(X_i) = \emptyset$.

For example, four complete terms $T_{11}$, $T_{12}$, $T_{15}$, and $T_{16}$ appear in Table 4 because there are two nonzero coordinates in $T_{11} = (0, 1, 0, 1)$, which can be discarded without further consideration in calculating the network reliability to save on runtime and reduce memory. If $T_{16}$ is the only complete term, $T_{16}$ must be included in the final reliability.

Hence, if the final IET term (that is, each coordinate of its corresponding vector has a value of one) is the only complete term, it must be included in the network reliability calculation; otherwise, all complete terms can be removed directly.

## 5. PROPOSED BAT-IET

Based on the four major novelties discussed in Section 4, the proposed BAT-IET pseudo-code, an example, and comparisons with other methods are presented in this section.

### 5.1 Pseudo-Code

The proposed RIE pseudo-code is presented below.

**Algorithm for proposed RIE**

**Input:** A heterogeneous-arc binary-state network $G(V, E, \mathbf{D})$, source node 1, sink node $n$, and all undirected MPs: $Q_1, Q_2, \ldots, Q_p$ act as input.

**Output:** $R(G)$.

**STEP 0.** Transfer each undirected MP to a directed MP based on Section 4.1.

**STEP 1.** Let $i = 2, j = 3, R = 0, T_1 = \emptyset, T_2 = P_1, S(T_1) = -1, S(T_2) = 1, \Pr(T_1) = 1$, and
$\Pr(T_2) = \prod \Pr(a)$ for all $a$ with $V^*(T_2(a)) \neq 0$.

**STEP 2.** Let $k = 1$.

**STEP 3.** Let $V^*(T_j) = V^*(T_k \cap P_i)$ based on Section 4.4.

**STEP 4.** If $T_j$ is a complete term, discard $T_j$ based on Section 4.5 and go to STEP 8.

**STEP 5.** Let $\Pr(T_j) = \Pr(T_k) \times \prod \Pr(a)$ for all $a$ with $V^*(T_k(a)) \neq V^*(T_i(a))$ based on Section 4.4.

**STEP 6.** If $S(T_k) = -1$, let $S(T_j) = 1$, $R = R + \Pr(V^*(T_j))$, $j = j + 1$, and go to STEP 8.

**STEP 7.** Let $S(T_j) = -1$, $j = j + 1$, and $R = R - \Pr(V^*(T_j))$.

**STEP 8.** If $k < p^*$, let $k = k + 1$ and go to STEP 3.

**STEP 9.** If $i < p$, let $p^* = j$, $i = i + 1$, and go to STEP 2. Otherwise, $R = R(G)$ and halt.

STEP 0 transfers all undirected MPs to directed MPs based on the procedure proposed in Section 4.1. The loop from STEPs 1 to 8 is based on Section 4.3. STEPs 3 and 4 are revised STEPs R2 and R3 in Section 4.3, based on Sections 4.4 and 4.5, respectively. The time complexity of the proposed RIE is at least equal to that of the recursive BAT-based IET, which is $O(n2^p)$.

**5.2 Step-by-Step Example**

Calculating the exact binary-state network reliability is both NP-hard and #P-hard [48, 49], which implies that the difficulty increases exponentially with the problem size. Only four undirected MPs exist in Figure 1: $\{a_1, a_4\}$, $\{a_1, a_3, a_5\}$, $\{a_2, a_5\}$, and $\{a_2, a_3, a_4\}$. Hence, to allow readers to understand the proposed RIE immediately, Figure 1 is exemplified to explain the proposed RIE step by step, as follows:

**STEP 0.** From Section 4.1, these four undirected MPs are transferred to directed MPs as follows: $P_1 = \{e_{1,2}, e_{2,4}\}$, $P_2 = \{e_{1,2}, e_{2,3}, e_{3,4}\}$, $P_3 = \{e_{1,3}, e_{3,4}\}$, and $P_4 = \{e_{1,3}, e_{3,2}, e_{2,4}\}$.

**STEP 1.** Let $i = 2, j = 3, R = 0, T_1 = \emptyset, T_2 = P_1 = \{e_{1,2}, e_{2,4}\}, S(T_1) = -1, S(T_2) = 1$, $\Pr(T_1) = 1$, and $\Pr(T_2) = \prod_a \Pr(a) = \Pr(e_{1,2}) \times \Pr(e_{2,4}) = 0.81$, where for all $a$ with $V^*(T_2(a)) \neq 0$.

**STEP 2.** Let $k = 1$.

**STEP 3.** Let $V^*(T_3) = V^*(T_1 \cap P_2) = (1, 0, 1, 0, 1)$ because $V^*(T_1) = (0, 0, 0, 0, 0)$ and $V^*(P_2) = (1, 0, 1, 0, 1)$ based on Section 4.4.

**STEP 4.** $T_3$ is not a complete term.

**STEP 5.** Let $\Pr(T_3) = \Pr(T_1) \times \prod \Pr(a) = 1 \times \Pr(v_{1,2}=1) \times \Pr(v_{2,3}=1) \times \Pr(v_{3,4}=1) = 0.729$, where for all $a \in \{e_{1,2}, e_{2,3}, e_{3,4}\}$ with $V^*(T_1(a)) \neq V^*(T_3(a))$.

**STEP 6.** Because $S(T_1) = -1$, let $T_3 = 1$, $R = R + \Pr(V^*(T_3)) = 0.81$, $j = j + 1 = 4$, and go to STEP 8.

**STEP 8.** Because $k = 1 < p^* = 2$, let $k = k + 1 = 2$ and go to STEP 3.

**STEP 3.** Let $V^*(T_4) = V^*(T_2 \cap P_2) = (1, 0, 1, 1, 1)$ because $V^*(T_2) = (1, 0, 0, 1, 0)$ and $V^*(P_2) = (1, 0, 1, 0, 1)$.

**STEP 4.** $T_4$ is not a complete term.

**STEP 5.** Let $\Pr(T_4) = \Pr(T_2) \times \prod \Pr(a) = 0.81 \times \Pr(v_{2,3}=1) \times \Pr(v_{3,4}=1) = 0.6561$, where for all $a \in \{e_{2,3}, e_{3,4}\}$ with $V^*(T_2(a)) \neq V^*(T_4(a))$.

$$\vdots$$

**STEP 8.** Because $k = 2 < p^* = 8$, let $k = k + 1 = 3$ and go to STEP 3.

**STEP 3.** Let $V^*(T_{11}) = V^*(T_3 \cap P_4) = (1, 1, 3, 1, 1)$ because $V^*(T_3) = (1, 0, 1, 0, 1)$ and $V^*(P_4) = (0, 1, 2, 1, 0)$.

**STEP 4.** Because $T_{11}$ is a complete term, discard $T_{11}$ and go to STEP 8.

$$\vdots$$

The complete procedure and the results that were obtained from the proposed RIE are presented in Table 5.

**Table 5.** Complete results for Figure 1 from the RIE.

| $i$ | $X_i$ | $T_i$ | $V^*(T_i)$ | $\Pr(T_i)$ | $S(T_i)$ | $R$ |
|---|---|---|---|---|---|---|
| 1 | (0, 0, 0, 0) | $T_1$ | (0, 0, 0, 0, 0) | 1 | −1 | |
| 2 | (1, 0, 0, 0) | $T_2=\{e_{1,2}, e_{2,4}\}$ | (1, 0, 0, 1, 0) | 0.81=0.9×0.9 | 1 | 0.81 |
| 3 | (0, 1, 0, 0) $T_1 \cap P_2 = \{e_{1,2}, e_{2,3}, e_{3,4}\}$ | (1, 0, 1, 0, 1) | 0.729=1×(0.9×0.9×0.9) | 1 | 1.539 |
| 4 | (1, 1, 0, 0) $T_2 \cap P_2 = \{e_{1,2}, e_{2,4}, e_{2,3}, e_{3,4}\}$ | (1, 0, 1, 1, 1) | 0.6561=0.81×(0.9×0.9) | −1 | 0.8829 |
| 5 | (0, 0, 1, 0) $T_1 \cap P_3 = \{e_{1,3}, e_{3,4}\}$ | (0, 1, 0, 0, 1) | 0.81=1×(0.9×0.9) | 1 | 1.6929 |
| 6 | (1, 0, 1, 0) $T_2 \cap P_3 = \{e_{1,2}, e_{2,4}, e_{1,3}, e_{3,4}\}$ | (1, 1, 0, 1, 1) | 0.6561=0.81×(0.9×0.9) | −1 | 1.0368 |
| 7 | (0, 1, 1, 0) $T_3 \cap P_3 = \{e_{1,2}, e_{2,3}, e_{3,4}, e_{1,3}\}$ | (1, 1, 1, 0, 1) | 0.6561=0.729×(0.9) | −1 | 0.3807 |
| 8 | (1, 1, 1, 0) $T_4 \cap P_3 = \{e_{1,2}, e_{2,4}, e_{2,3}, e_{3,4}, e_{1,3}\}$ | (1, 1, 1, 1, 1) | 0.59049=0.6561×(0.9) | 1 | 0.97119 |
| 9 | (0, 0, 0, 1) $T_1 \cap P_4 = \{e_{1,3}, e_{3,2}, e_{2,4}\}$ | (0, 1, 2, 1, 0) | 0.648=1×(0.9×0.8×0.9) | 1 | 1.61919 |
| 10 | (1, 0, 0, 1) $T_2 \cap P_4 = \{e_{1,2}, e_{2,4}, e_{1,3}, e_{3,2}\}$ | (1, 1, 2, 1, 0) | 0.5832=0.81×(0.9×0.8) | −1 | 1.03599 |
| 11 | (0, 0, 1, 1) $T_5 \cap P_4 = \{e_{1,3}, e_{3,4}, e_{3,2}, e_{2,4}\}$ | (0, 1, 2, 1, 1) | 0.5832=0.81×(0.8×0.9) | −1 | 0.45279 |
| 12 | (1, 0, 1, 1) $T_6 \cap P_4 = \{e_{1,2}, e_{2,4}, e_{1,3}, e_{3,4}, e_{3,2}\}$ | (1, 1, 3, 1, 1) | 0.52488=0.6561×(0.8) | 1 | 0.97767 |

## 5.3 Computation Experiments

Most new algorithms are tested on benchmark graphs, as illustrated in Figure 3 [34, 35, 44, 48, 49], to validate their effectiveness and efficiency in calculating network reliability problems. Hence, the superiority of the proposed RIE is also demonstrated and verified on these 11 binary-state benchmark networks by considering that all arcs are heterogeneous, such that $\Pr(e_{i,j}) = \begin{cases} 0.9 & \text{if } i < j \\ 0.8 & \text{otherwise} \end{cases}$ in the experiment on heterogeneous-arc binary network problems.

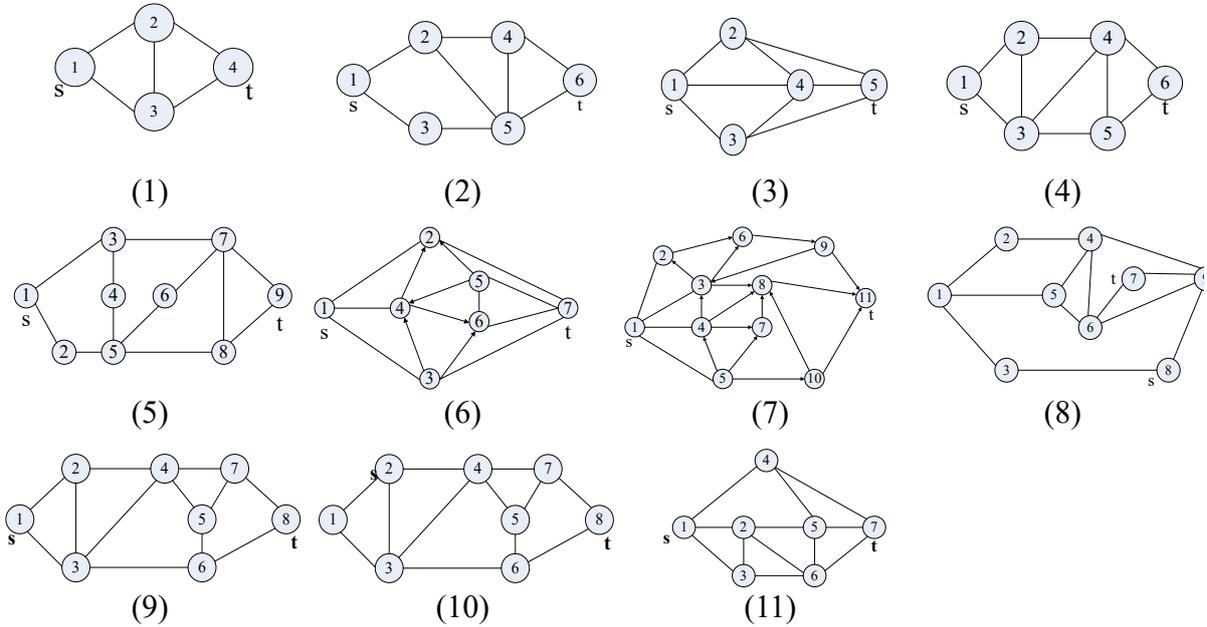

**Figure 3.** 11 benchmark binary-state networks used in the test

The results of the proposed RIE are compared with those of QIE and QIE* for calculating the reliability of binary-state networks because QIE is still the most efficient IET [35]. As in most current algorithms [34, 35, 44, 48, 49], each undirected arc in QIE is replaced with two directed arcs. QIE* is a special QIE that uses the concept of the proposed augmented-state vectors; that is, it is not necessary to transfer each heterogeneous arc into two directed arcs. Note that both the time complexity and runtime are squares of the original values after replacing each undirected arc with two directed arcs.

Moreover, for a fair evaluation, the proposed RIE, QIE, and QIE* are tested in the same test environments, as indicated in Table 6.

Table 6. Computer Environments and problem settings.

| Item | Environment |
|---|---|
| Operation System | 64-bit Windows 10 |
| Compiler | DEV C$^{++}$ 5.11 |
| CPU | Intel Core i7-6650U @ 2.20GHz 2.21GHz |
| RAM | 32 GB |
| Stopping Criteria | 10 hours |

A comparison of the results is presented in Table 7. The best results among the RIE, QIE, and QIE* are denoted in bold. The notations $n$, $m$, $m^*$, $p$, and $R$ are defined in Section 2.2.

Table 7. Comparison of the RIE and QIE [10].

| Fig. | $n$ | $m$ | $m^*$ | $p$ | $R$ | $N_{QIE}$ | $T_{QIE}$ | $N_{RIE}$ | $T_{RIE}$ |
|---|---|---|---|---|---|---|---|---|---|
| 1 | 4 | 5 | 6 | 4 | 0.9776700000 | 15 | 0 | **11** | 0 |
| 2 | 6 | 8 | 12 | 7 | 0.9673611300 | 127 | 0 | **107** | 0 |
| 3 | 5 | 8 | 10 | 9 | 0.9974331900 | 511 | 0.001 | **339** | **0** |
| 4 | 6 | 9 | 14 | 13 | 0.9769992390 | 8191 | 0.001 | **6171** | **0** |
| 5 | 9 | 12 | 20 | 13 | 0.9643784272 | 8191 | 0 | **6171** | 0 |
| 6 | 7 | 14 | 19 | 14 | 0.9963554556 | 16383 | 0.001 | 16383 | **0** |
| 7 | 11 | 21 | 26 | 18 | 0.9939922430 | 262143 | 0.009 | 262143 | **0.008** |
| 8 | 9 | 13 | 22 | 18 | 0.9593959133 | 262143 | 0.007 | 262143 | **0.005** |
| 9 | 8 | 12 | 20 | 24 | 0.9747748172 | 16777215 | 0.354 | **8573035** | **0.22** |
| 10 | 8 | 12 | 19 | 20 | 0.9838074640 | 1048575 | 0.024 | **550127** | **0.014** |
| 11 | 7 | 12 | 18 | 25 | 0.9971989718 | 33554431 | 0.758 | **18182251** | **0.443** |

According to Table 7, the number of obtained IET terms and runtimes all increase exponentially with the number of MPs, and these meet the NP-hard and #P-hard characteristics of network reliability [48, 49]. These observations indicate that RIE and QIE$^*$ are more efficient than QIE, owing to the benefit of using the proposed augmented-state vectors. Furthermore, both $T_{QIE^*}$ and $T_{RIE}$ are zero when the number of MPs is $\leq 13$ (e.g., Figure 3(1) to 3(5)); that is, both RIE and QIE$^*$ can easily solve small-scale problems.

The proposed augmented vector implies that the concept of directed MPs is also included. Hence, the above result confirms the performance of the proposed augmented-state vectors and directed MPs in decreasing the runtime.

To investigate the effects of the other proposed concepts, including the recursive concept and complete terms, we compare the proposed RIE with QIE$^*$, which implements the concepts of the proposed augmented-state vectors and undirected MP in QIE in this experiment on heterogeneous-arc binary network problems.

For the medium-sized problem, the proposed RIE starts to outperform QIE$^*$ in all aspects, including the number of obtained terms and runtimes. This is because RIE can detect and discard complete terms as soon as they are identified and does not need to involve these terms for further considerations to save time. We always obtain $N_{RIE} \leq N_{QIE^*}$ and the difference becomes extraordinary after $p$ reaches its maximum value, as indicated in Table 7.

According to Table 7, the runtime tends to be proportional to the values of the obtained terms. Hence, RIE is more efficient than QIE$^*$ owing to $N_{RIE} < N_{QIE^*}$. Therefore, the above results validate that the concept of the proposed complete terms can reduce the runtime and IET terms.

For $N_{RIE} = N_{QIE^*}$, RIE is still more efficient than QIE$^*$ because the proposed recursive concept reduces the number of comparisons and calculations in the intersections and the probability of each term.

The foregoing comparisons of the experimental results demonstrate the superiority of the RIE over both QIE and QIE[*], which also implement the proposed directed MPs and augmented-state vectors. The time complexity presented in Section 5.1 is also confirmed by the relationship between $T_{RIE}$ and $N_{RIE}$ in the results.

## 6. CONCLUSIONS

Network reliability reflects the probability of the current status of a network and is an essential indicator for evaluating the network performance. The heterogeneous-arc network reliability problem is critical in real-life applications. However, its runtime is the square of the network without heterogeneous arcs, and it has not attracted substantial attention in research.

Conditional network reliability problems always depend on MPs, and IET is the most convenient tool for calculating the final reliability after obtaining all MPs. Hence, a new, simple, and efficient IET known as RIE has been proposed to solve the heterogeneous-arc network reliability problem after determining all MPs.

RIE is the first BAT-based recursive IET that focuses on the heterogeneous-arc network reliability problem. It integrates the BAT, directed MPs, recursive concepts, augmented-state vectors, and complete terms. The BAT provides the proposed recursive concept for the implementation of IET to reduce the number of intersections by at least $(1/p)$ of the original IET, which can reduce the number of directed arcs by half, whereas the concept of the complete terms can decrease the number of IET terms.

With the implementation of the proposed RIE, the time complexity and runtime for solving the heterogeneous-arc network reliability problem are improved to that of the homogeneous-arc network reliability.

To ascertain the advantages, in future work, the proposed RIE will undergo fairly advanced comparisons with other known related algorithms; for example, the sum-of-disjoint product algorithms and binary decision diagram algorithms, based on more open-source datasets.

The memory overflow problem is a major drawback of all recursive algorithms, including recursive SDP [50] and RIE. For example, RIE runs out of memory if the problem size is too large, because all terms must be saved to guarantee that each IET term can be obtained from the previous two IET terms. Moreover, a recursive algorithm is always less efficient than a direct version of the algorithm [34] and lacks an efficient means of managing the required memory space. Hence, the direct version of RIE is another important future investigation.

## ACKNOWLEDGEMENTS

This research was supported in part by the Ministry of Science and Technology, R.O.C. under grant MOST 107-2221-E-007-072-MY3 and MOST 110-2221-E-007-107-MY3. This article was once submitted to arXiv as a temporary submission that was just for reference and did not provide the copyright.

## REFERENCES

[1] Zhang X, Huang Y, Li L, Yeh WC. Power and Capacity Consensus Tracking of Distributed Battery Storage Systems in Modular Microgrids. Energies 2018; 11 (6): 1439.

[2] Chen P, Li Y, Wang K, Zuo MJ, Heyns PS, Baggeröhr S. A threshold self-setting condition monitoring scheme for wind turbine generator bearings based on deep convolutional generative adversarial networks. Measurement 2021; 167: 108234.

[3] Anchieta TFF, Santos SAR, Brentan BM, Carpitella S, Izquierdo J. Managing expert knowledge in water network expansion project implementation. IFAC-PapersOnLine 2021; 54 (17): 36–40.

[4]   Cui Z, Chen J, Liu C, Zhao H. Time-domain continuous power flow calculation of electricity–gas integrated energy system considering the dynamic process of gas network. Energy Reports 2022; 8 (5): 597–605.

[5]   Zhou J, Coit DW, Felder FA, Wang D. Resiliency-based restoration optimization for dependent network systems against cascading failures. Reliability Engineering & System Safety 2021; 207: 107383.

[6]   Lin YK, Huang CF, Liao YC, Yeh CC. System reliability for a multistate intermodal logistics network with time windows. International Journal of Production Research 2017; 55 (7): 1957–1969.

[7]   Yeh WC, Lai CM, Tsai JY. Simplified swarm optimization for optimal deployment of fog computing system of industry 4.0 smart factory. Journal of Physics Conference Series 2019; 1411: 012005.

[8]   Dogra A, Jha RK, Jain S. A survey on beyond 5G network with the advent of 6G: Architecture and emerging technologie. IEEE Access 2020; 9: 67512–67547.

[9]   Yeh WC, Lin JS. New parallel swarm algorithm for smart sensor systems redundancy allocation problems in the Internet of Things. The Journal of Supercomputing 2018; 74 (9): 4358–4384.

[10]  Kakadia D, Ramirez-Marquez JE. Quantitative approaches for optimization of user experience based on network resilience for wireless service provider networks. Reliability Engineering & System Safety 2020; 193: 106606.

[11]  Abbasi B, Hosseinifard SZ, Coit DW. A neural network applied to estimate Burr XII distribution parameters. Reliability Engineering & System Safety 2010; 95 (6): 647–654.

[12]  Chen P. Li Y, Wang K, Zuo MJ. An automatic speed adaption neural network model for planetary gearbox fault diagnosis. Measurement 2021: 171: 108784.

[13]  Chen YF, Lin YK, Huang CF. Using Deep Neural Networks to Evaluate the System Reliability of Manufacturing Networks. International Journal of Performability Engineering 2021; 17 (7).

[14]  Yeh WC. A squeezed artificial neural network for the symbolic network reliability functions of binary-state networks. IEEE transactions on neural networks and learning systems 2017; 28 (11): 2822–2825.

[15]  Zhu W, Yeh W, Cao L, Zhu Z, Chen D, Chen J, Li A, Lin Y. Faster Evolutionary Convolutional Neural Networks Based on iSSO for Lesion Recognition in

Medical Images. BASIC & CLINICAL PHARMACOLOGY & TOXICOLOGY 2019; 124: 329–329.

[16] Levitin G. Reliability of acyclic multi-state networks with constant transmission characteristics of lines. Reliability Engineering & System Safety 2002; 78 (3): 297–305.

[17] Niu YF, Song YF, Xu XZ, Zhao X. Efficient reliability computation of a multi-state flow network with cost constraint. Reliability Engineering & System Safety 2022; 222: 108393.

[18] Yeh WC. A Novel Method for the Network Reliability in terms of Capacitated-Minimal-Paths without Knowing Minimal-Paths in Advance. Journal of the Operational Research Society 2005; 56(10): 1235–1240.

[19] Yeh WC. A Simple Minimal Path Method for Estimating the Weighted Multi-Commodity Multistate Unreliable Networks Reliability. Reliability Engineering & System Safety 2008; 93(1): 125–136.

[20] Hao Z, Yeh WC, Zuo M, Wang J. Multi-distribution multi-commodity multistate flow network model and its reliability evaluation algorithm. Reliability Engineering & System Safety 2020; 193: 106668.

[21] Levitin G. A universal generating function approach for the analysis of multi-state systems with dependent elements. Reliability Engineering & System Safety 2004; 84(3): 285–292.

[22] Yeh WC. A squeezed artificial neural network for the symbolic network reliability functions of binary-state networks. IEEE transactions on neural networks and learning systems 2016; 28: 2822–2825.

[23] Sanseverino CMR, Ramirez-Marquez JE. Uncertainty propagation and sensitivity analysis in system reliability assessment via unscented transformation. Reliability Engineering & System Safety 2014; 132: 176–185.

[24] Lin YK. Extend the quickest path problem to the system reliability evaluation for a stochastic-flow network. Computers & Operations Research 2003; 30(4): 567–575.

[25] Yeh WC, Khadiri ME. A New Universal Generating Function Method for Solving the Single -Quick-Path Problem in Multistate Flow Networks. IEEE Transactions on Systems Man and Cybernetics - Part A Systems and Humans 2012; 42(6): 1476–1484.

[26] Hao Z, Yeh WC, Tan SY. One-batch Preempt Deterioration-effect Multi-state Multi-rework Network Reliability Problem and Algorithms. Reliability Engineering and System Safety 2021; 215: doi.org/10.1016/j.ress.2021.107883.

[27] Yeh WC, Lin YC, Chung YY. Performance analysis of cellular automata Monte Carlo Simulation for estimating network reliability. Expert Systems with Applications 2010; 37(5): 3537–3544.

[28] Niu YF, Shao FM. A practical bounding algorithm for computing two-terminal reliability based on decomposition technique. Computers & Mathematics with Applications 2011; 61(8): 2241–2246.

[29] Forghani-elahabad M, Kagan N, Mahdavi-Amiri N. An MP-based approximation algorithm on reliability evaluation of multistate flow networks. Reliability Engineering & System Safety 2019; 191: 106566.

[30] Yeh WC. A path-based algorithm for evaluating the k-out-of-n flow network reliability. Reliability Engineering & System Safety 2005; 87 (2): 243–251.

[31] Forghani-elahabad M, Kagan N. Reliability evaluation of a stochastic-flow network in terms of minimal paths with budget constraint. IISE Transactions 2019; 51(5): 547–558.

[32] Yeh WC. An Improved Method for the Multistate Flow Network Reliability with Unreliable Nodes and the Budget Constraint Based on Path Set. IEEE Transactions on Systems, Man, and Cybernetics: Systems 2011; 41(2): 350–355.

[33] Yeh WC, Tan SY. Simplified Swarm Optimization for the Heterogeneous Fleet Vehicle Routing Problem with Time-Varying Continuous Speed Function. Electronics 2021; 10: doi:10.3390/electronics10151775.

[34] Yeh WC. Novel Binary-Addition Tree Algorithm (BAT) for Binary-State Network Reliability Problem. Reliability Engineering and System Safety 2020; 208: 107448.

[35] Hao Z, Yeh WC, Wang J, Wang GG, Sun B. A quick inclusion-exclusion technique. Information Sciences 2019; 486: 20–30.

[36] Yeh WC. A Simple Heuristic Algorithm for Generating All Minimal Paths. IEEE Transactions on Reliability 2007; 56(3): 488–494.

[37] Levitin G. The universal generating function in reliability analysis and optimization, Springer, 2005.

[38] Yeh WC. A novel node-based sequential implicit enumeration method for finding all d-MPs in a multistate flow network. Information Sciences 2015; 297: 283–292.

[39] Kessler D, Schiff J. Inclusion-Exclusion Redux. Electronic Communications in Probability 2002; 7: 85–96.

[40] Dohmen K. Improved inclusion-exclusion identities via closure operators. Discrete Math. Theoret. Comput. Sci. 2000; 4: 61–66.

[41] Dohmen K. Inclusion-Exclusion and Network Reliability. The Electronic Journal of Combinatorics 1998; 5(1): 36, Internet: http://www.combinatorics.org.

[42] Shen Y. A new simple algorithm for enumerating all minimal paths and cuts of a graph. Microelectronics Reliability 1995; 35(6): 973–976.

[43] Yeh WC. Search for minimal paths in modified networks. Reliability Engineering & System Safety 2002; 75 (3): 389–395.

[44] Yeh WC. A Quick BAT for Evaluating the Reliability of Binary-State Networks. Reliability Engineering and System Safety 2020; 216: doi.org/10.1016/j.ress.2021.107917.

[45] Su YZ, Yeh WC. Binary-Addition Tree Algorithm-Based Resilience Assessment for Binary-State Network Problems. Electronics 2020; 9(8): 1207.

[46] Yeh WC, Kuo CC. Predicting and Modeling Wildfire Propagation Areas with BAT and Maximum-State PageRank. Appl. Sci. 2020; 10: 8349. https://doi.org/10.3390/app10238349.

[47] Yeh WC, Lin E, Huang CL. Predicting Spread Probability of Learning-Effect Computer Virus. Complexity 2021; 2021: doi.org/10.1155/2021/6672630.

[48] Shier D, Network Reliability and Algebraic Structures, Clarendon Press, New York, NY, USA, 1991.

[49] Colbourn CJ, The combinatorics of network reliability, Oxford University Press, New York, 1987.

[50] Zuo MJ, Tian Z, Huang HZ, An efficient method for reliability evaluation of multistate networks given all minimal path vectors, IIE Transactions 2007, 39(8):811-817.

[51] https://sites.google.com/view/wcyeh/source-code-download